\documentclass{article}
\newcommand{\ket}[1]{{|#1\rangle}} 

\author{Christof Zalka}
\title{Shor's algorithm with fewer (pure) qubits}

\begin{document}
\maketitle

\begin{abstract}
  In this note we consider optimised circuits for implementing Shor's
  quantum factoring algorithm. First I give a circuit for
  which non of the about $2 n$ qubits need to be initialised (though
  we still have to make the usual $2 n$ measurements later on).
  % The modification makes the algorithm a bit slower, but this can be
  % compensated by using a ``trinary'' quantum Fourier transformation, thus
  % one of order $3^{n'}$, instead of the usual binary one.  A second
  % observation may be of more practical interest.
  Then I show how the modular additions in the algorithm can be
  carried out with a superposition of an arithmetic sequence. This
  makes parallelisation of Shor's algorithm easier. Finally I show how
  one can factor with only about $1.5n$ qubits, and maybe even fewer.
  % The presentation is sketchy, also in the hope of making it more didactic
  % and readable.
\end{abstract}

\section{Introduction}
Over the years people have looked into how Shor's quantum factoring
algorithm could actually be implemented on a quantum computer. In
particular people have tried to optimise the algorithm (the quantum
part of it). An early work was by C\'esar Miquel in 1996
\cite{miquel}. Later, people were mostly trying to reduce the size of
a quantum computer needed for factoring (thus the number of
qubits). The circuit of St\'ephane Beauregard from 2003 \cite{beau}
uses only $2 n+3$ qubits to factor an $n$-bit number.

An important observation has always been that much of the computation can be
done classically. A lot of quantities can be ``precomputed'' conventionally.
% classically and the output of the quantum algorithm of course also needs
% ``postprocessing'' to factor the number (with good probability).
For the most part the quantum circuits are just classical reversible circuits
to compute a modular exponentiation, which is reduced to many (actually some
$4 n^2$) modular additions of $n$-bit numbers.  A non-classical circuit is Tom
Draper's addition technique from 2000 \cite{draper} using a quantum Fourier
transform which allows to do without the additional $n$ work qubits which are
usually needed to hold the carry bits.

The first observation in this context which I present is more of
theoretical than of practical interest. Parker and Plenio
(\cite{plenio}, 2000) have pointed out that $n$ of the $2 n$ qubits
used in these implementations need not be initialised (set to
$\ket{0}$) at the beginning. (But in the title and abstract of their
paper they seem to claim even more than that). I now show how the
algorithm can be modified so that actually none of the about $2 n$
qubits need to be initialised. Still, in mine as in their algorithm we
have to do some $2 n$ measurements in the course of the algorithm. As
measuring a qubit also resets it, it would therefore not be correct to
say that we need no pure qubits at all. The technique works by
modifying the (controlled) modular multiplications. Instead of the
usual modular multiplication $\ket{\alpha,0} \to \ket{a \cdot
\alpha,0}$, (where $a$ is a classically known number) I show how to do
$\ket{\alpha,\beta} \to \ket{a \cdot \alpha,a^{-1} \cdot \beta}$
without need for further work qubits.  Instead of the usual 2 modular
multiplications for this step (one by $a$, the other by $a^{-1}$,
everything $\bmod ~N$), I now need 3 of them. In a further observation
I show how the slowdown caused by this can be compensated by using a
``trinary'' quantum Fourier transform.

% (actually even a bit overcompensated). In each step, depending on some
% ``control-qubits'', we can multiply with $a$, do nothing, or indeed we can,
% without much more work, also multiply with $a^{-1}$. This suggests using a
% ``trinary'' quantum fast Fourier transform instead of the usual binary one,
% reducing the runtime by a factor of $\log_2 3 \approx 1.58 \dots$.

The second observation is of a different kind. It looks at the modular
additions (modulo the number $N$ to be factored) out of which the algorithm
essentially consists. The fact that these are not usual additions, but modular
ones causes some complications (effectively several additions have to be
made). In particular when using Draper's ``Fourier-addition'' we need to
Fourier transform back to the usual basis after each addition, which is
costly. I now propose an approximate way to compute modular additions with a
quantum method (as opposed to a classical reversible circuit). We need to make
the $n$-qubit registers larger by some O($\log n$) qubits. Then instead of a
number $\ket{b}$ with $0\le b < N$, we consider an equal (``uniform'')
superposition $\sum_x \ket{b+x N}$.  Using this technique allows to simplify
the overall outline of the algorithm and may allow a considerable speedup,
mainly because it allows a lot of parallelisation. The depth of the overall
circuit for Shor's algorithm can then be reduced to O($n^2$).

The third result gives a sizable reduction in the number of qubits
needed in Shor's algorithm. Namely from the present number of about
$2n$ qubits (Beauregard) to about $1.5n$. I show two things. One is
how to replace modular multiplication by an $n$-bit factor with two
multiplications by factors half this length. Then I show how such
``short'' multiplications can be done with accordingly fewer
qubits. (If somehow we manage to further factor these multipliers,
further reductions down to closer to $1.0n$ would be possible.) The
presentation, especially for this third result, is sketchy.

\section{Review of circuits for Shor's algorithm}

The number $N$ we want to factor is an $n$-bit number, thus $n=\lceil
\log_2 N \rceil$ ($\log_2 N$ rounded up). $N$ may e.g. be the product
of two unknown large prime numbers, thus $N=p \cdot q$. The problem of
factorisation can be reduced to finding the period of the periodic
function $f(x)= a^x \bmod N$, where $a$ is essentially an arbitrary
integer (really $0<a<N$ and $a$ should be coprime with $N$). The
``order finding'' quantum algorithm to do this can be described as
follows. We have a $2 n$-qubit register and an $n$-qubit register:
$$ \sum_{x=0}^{2^{2n}-1} \ket{x} \ket{0} \quad \to \quad
\sum_{x=0}^{2^{2n}-1} \ket{x} \ket{a^x \bmod N} \quad \to \quad
\sum_{x=0}^{2^{2n}-1} FT(\ket{x})~ \ket{a^x \bmod N}
$$
After Fourier transforming the first ($2n$-qubit) register, we measure it.
(From the result, the period of $~a^x \bmod N~$ can be computed classically
with good success probability.) The modular exponentiation can be computed by
noting (everything is understood to be $\bmod N$):
$$ a^{x=x_0+2 x_1 + \dots +2^{2n-1} x_{2n-1}} \quad = \quad
 a^{x_0} \cdot {\left(a^2\right)}^{x_1}
\cdot {\left(a^{2^2}\right)}^{x_2}
\cdot ~\dots~ \cdot {\left(a^{2^{2n-1}}\right)}^{x_{2n-1}} $$
where $x_0, x_1, \dots$ are the bits of $x$. The numbers $a^{2^i}$ can
be computed through repeated squaring (at every step reducing $\bmod
~N$). Thus in the end we really only have to make a sequence of
(modular) multiplications, each conditioned on a different bit of
$x$. Now look at the subsequent Fourier transform and measurement of
the $x$-register. Fourier transform and measurement can be combined
(``Fourier sampling'', ``semiclassical Fourier transform'') so that we
first measure the high bits of the $x$-register (one after the other,
starting from the highest value one). Indeed the whole procedure can
be described as simply measuring one qubit after another, each in a
basis depending on the previous measurement outcomes (of the higher
qubits). Given that the $x$-register is initially in a product state,
we see that we need not keep these $2n$ qubits all at the same time.
For each $x$-qubit we can prepare it in state $1/\sqrt{2}
(\ket{0}+\ket{1})$, then use it to control a modular multiplication
and then measure it in the appropriate basis. (Note that for this to
work, we have to carry out the multiplications with $a^{2^i}$ in
``reversed order'', thus $i=2n-1, \dots ,1,0$.) Thus instead of the
whole $2n$-qubit $x$-register, now we need only 1 qubit.

Each modular multiplication is of the form $\ket{\alpha} \to \ket{a
\cdot \alpha}$, where $\alpha$ is the number initially in the
register, and we simply use ``$a$'' for any number we want to multiply
with (really it's a power of $a$).  Parker and Plenio \cite{plenio}
have pointed out that if we don't initialise the register, we
effectively compute $~\alpha \cdot a^x \bmod N~$ instead of $~a^x
\bmod N~$, where $\alpha$ is the number that happens to be initially
in the register.  (Why we can think of ``not initialised'' in this way
is explained later.) For most $\alpha$ this function (of $x$) has the
same periodicity as the original one (namely for all $\alpha$ that are
coprime with $N$). So Parker and Plenio said that none of the initial
qubits needed to be initialised. For this they assumed that they had a
unitary operator that would modularly multiply a register with any
given $a$ coprime with $N$ (indeed this operation is reversible and
thus unitary): $U_a: \ket{\alpha} \to \ket{a \cdot \alpha}$. But we
don't know how to do this efficiently without a supply of another
O($n$) properly initialised qubits. In the usual technique we use an
auxilliary $n$-qubit register and make two steps whereby we add a
multiple of one register to the other register (everything $\bmod N$):
$$\ket{\alpha,0} \quad \to \quad \ket{\alpha,a \cdot \alpha} \quad \to
\quad \ket{\alpha - a^{-1} \cdot (a \cdot \alpha),a \cdot \alpha} =
\ket{0,a \cdot \alpha} $$
%\quad \to \quad \ket{a \cdot \alpha,0} $$
%
where in the second step we use (modular) multiplication with $a^{-1}
(\bmod N)$ (which we can precompute classically using Euclid's
algorithm).
%The last step is simply a swap.

\section{Not initialising any qubits} 

The above sequence of operations doesn't give the desired result if
the auxilliary register is not initialised to $\ket{0}$, but this can
easily be fixed. We start by applying the usual two steps to registers
in any initial state but add an appropriate third step:
$$\ket{\alpha,\beta} ~\to~ \ket{\alpha,\beta +a \alpha} ~\to~
\ket{\alpha - a^{-1} (\beta +a \alpha)=-a^{-1} \beta,\beta +a \alpha} ~\to~ 
\ket{-a^{-1} \beta,a \alpha} $$
Thus by adding a SWAP and multiplication by $-1$ (which is rather easy), we
can do $\ket{\alpha,\beta} \to \ket{a \cdot \alpha,a^{-1} \cdot \beta}$. This
is just as good for order finding, as the function $(a^x \alpha,a^{-x} \beta)$
has still the same period which we were looking for (at least for most numbers
$\alpha,\beta$). We can also write the sequence of operations as $2 \times 2$
matrices (acting on a pair of numbers in $Z_N$):
$$ 
\left( \begin{array}{cc} &+1 \\ -1& \end{array} \right) 
\left( \begin{array}{cc} 1& \\ a&1 \end{array} \right) 
\left( \begin{array}{cc} 1&-a^{-1} \\ &1 \end{array} \right) 
\left( \begin{array}{cc} 1& \\ a&1 \end{array} \right) 
\left( \begin{array}{c} \alpha \\ \beta \end{array} \right) =
\left( \begin{array}{c} a~ \alpha \\ a^{-1} \beta \end{array} \right)
$$
%
%Note that the basic operations which we use are represented by matrices with
%determinant 1. (More generally we can claim that by stringing together such
%operations we can generate any $2 \times 2$ matrix acting on $(Z_N)^2$ with
%determinant 1.)

The technique which was described here for modular multiplication actually
works in any group setting and can thus also be used for Shor's discrete
logarithm (dlog) algorithm.

Finally a remark on what it means to not initialise some
qubits. Clearly the algorithm may fail if e.g. someone has
intentionally prepared the qubits in some ``bad'' state. But we can
always ensure that they are in the maximally mixed state, namely by
randomly applying one of the four Pauli operators to each qubit. I
thus have simply assumed that ``uninitialised registers'' are in this
maximally mixed state, which is equivalent to thinking that the
register is at random in one of the ``computational'' basis states
$\ket{\alpha}$.

Actually so far we have only shown that the two main registers
involved in Shor's algorithm need not be initialised, but there are a
few more qubits, e.g. in Beauregards circuit we need 3 more qubits. As
this is a constant (and small) number, we still get a reasonable
success probability for Shor's algorithm if we don't initialise these
qubits, too. Thus one may claim that non of the initial qubits need to
be initialised.

\subsection{Saving time with a trinary quantum Fourier transform}

The multiplication steps are conditioned on bits of $x$, thus either
we multiply (modularly) with $a$, or we do nothing. But if we can do
$\ket{\alpha,\beta} \to \ket{a \alpha,a^{-1} \beta}$, it is clear that by
swapping the 2 registers before and after the operation, we can also do the
inverse: $\ket{\alpha,\beta} \to \ket{a^{-1} \alpha,a \beta}$. Note that such
controlled swaps are cheap compared to the multiplication steps. Thus
effectively we can multiply (say the first register) with either $a$, 1, or
with $a^{-1}$, depending on some control qubits. (Note that this ``control''
hardly increases the cost as the individual modular additions, out of which
the multiplications consist, are anyways conditional.) Thus is makes sense to
take advantage of all three possibilities. Multiplication with $a$, 1, or
$a^{-1}$ is very similar to multiplication with $a^2$, $a$, or $1$, it only
shifts the final (periodic) function by a constant amount: $f(x) \to
f(x-const.)$. Thus we now imagine that $x$ is given in ``trinary'' form
$x=x_0+3 x_1+3^2 x_2+ \dots +x_{2n'-1} 3^{2n'-1}$, where $n'$ is smaller than
$n$ by about a factor of $\log_2 3$. Note that it is not important whether we
store the bits of $x$ in actual physical ``qutrits'' or whether we e.g. use 2
qubits.

The final Fourier transform in Shor's algorithm now has to be replaced
with a trinary one (of order $3^{2n'}$). Everything works just as well
with such a trinary quantum Fourier transform, in particular the
``semiclassical'' version is analogous and we can essentially proceed
as before: Prepare a ``qutrit'' (possibly realised with 2 qubits) in
an equal superposition. Use it to control a modular multiplication
(now 3 possibilities!). Then measure the ``qutrit'' in a basis
determined by the previous measurement outcomes.

[For deriving the circuit for a trinary (or any ``$p$-ary'' of order
$p^n$) quantum Fourier transform and seeing that this works, I only
give some hints.  I find it useful to consider the transform of a
basis state $\ket{b} \to 1/\sqrt{\dots} \sum_x \omega^{bx}
\ket{x}$. It turns out that this is again a product state, where the
state of a given ``qudit'' depends only on some of the original digits
of $b$ (from $b_0$ up to a maximum). Start by seeing how the one qudit
which depends on all original digits can be obtained (namely by a
generalised Hadamard transform, followed by controlled phases). From
there on down, each qudit can be obtained in a similar way. Thus the
circuit has the same structure for any ``$p$-ary'' quantum Fourier
transform.]

The dominant cost in the controlled multiplications are the individual
multiplication steps. In the modified algorithm we have 3 instead of 2 of
these. But this slowdown is more than compensated through the use of the
trinary Fourier transform. Thus we now even use a bit less time by a factor
$3/2 \cdot \log_3 2 \approx 0.946 \dots$.

\section{Modular addition with equal ``coset superpositions''}

Most of Shor's algorithm actually consists of (conditional) modular additions
of a fixed (classically known) number to a quantum register: $\ket{\alpha} \to
\ket{\alpha+a \bmod N}$. This operation is of course reversible (just subtract
$a$ modulo N), but it can be a bit cumbersome to implement it. If we first
simply add and only then reduce modulo $N$ (thus possibly subtracting $N$),
then this last step would not be reversible. What one can do is to first make
a comparison (of the quantum register with a suitable ``classical'' number)
which determines whether $N$ will have to be subtracted. Then with the final
result in the quantum register, one can make another suitable comparison to
``uncompute'' this control qubit. Each of these comparisons essentially
amounts to an addition (or rather a subtraction). And, what is worse, when
using Draper's Fourier-addition technique, each time one has to Fourier
transform back to the usual basis for reading out the result of a comparison.

I now propose a modular addition technique which is faster, although it uses a
few more qubits. Instead of representing a number in $Z_N$ simply by a number
$b$ in the range $0\dots N-1$, we prepare an equal superposition of many terms
of the arithmetic sequence $b+x \cdot N$ with $x=0,1,\dots$. Thus we do
$\ket{b} \to \sum_x \ket{b+x N}$. It is enough if the new register is larger
by some $O(\log n)$ qubits (actually some $2 \log n+10$ qubits should be more
than enough for Shor's algorithm) and the range of $x$ will be accordingly.

Now we simply add (non-modularly) the number $a$ to this register:
$$ \sum_{x=0}^{x_{max}-1} \ket{b+x N} \quad \to \quad 
\sum_{x=0}^{x_{max}-1} \ket{a+b+x N} $$ 
The point is that the outcome is close to the (desired) outcome $\sum_x
\ket{(a+b) \bmod N +xN}$. Thus $\sum_x \ket{a+b+xN} \approx \sum_x \ket{(a+b)
  \bmod N +xN}$, in the sense that the fidelity (overlap) is close to
1. The ``ladder'' of peaks in the superposition simply may get shifted
by one period. So the loss of fidelity per addition is on the order of
$O(1/x_{max})$ and thus can be made very small.

\subsection{Converting to the ``coset representation''}

First let's see how we can transform back and forth between the usual
``representative element'' representation (of a number in $Z_N$) $\ket{b}$ and
the ``coset superposition'' representation $\sum_x \ket{b+xN}$. (Here the
coset is given by the elements $b+xN$ of the arithmetic sequence, the subgroup
being the multiples of $N$ and the overall group are the integers $Z$.) Given
$\ket{b}$, we need another small register prepared in an equal (``uniform'',
``flat'') superposition $1/\sqrt{x_{max}} \sum_x \ket{x}$.  It is easiest to
prepare this superposition if $x_{max}$ is a power of 2. (It may be
advantageous to choose a different $x_{max}$, and in this case, too it is not
difficult to prepare the equal superposition.) Then we apply a simple
(classical reversible) operation to these 2 registers. Thus:
$$ \mbox{prepare} \quad \frac{1}{\sqrt{x_{max}}} \sum_{x=0}^{x_{max}-1} 
\ket{x}, \qquad \mbox{then do} \quad \ket{b} \ket{x} \to \quad \ket{b+xN} $$
This last step is not hard to carry out. Imagine that we first extend the
range of the first (``$b$'') register (by adding a few qubits in state
$\ket{0}$ at the top end of the register). Then we go through the bits of $x$
from least significant to most significant. Conditioned on bit $x_i$ we add
$2^i N$ to the first register. Then we uncompute $x_i$ by checking whether the
content of the first register is $\ge 2^i N$.

\subsection{Use in Shor's algorithm}

First a quick review of how modular multiplications are decomposed into $n$
modular additions. (Actually this is similar to the decomposition of the
modular exponentiation into modular multiplications.) The individual modular
multiplication steps are of the form $\ket{\alpha} \ket{\beta} \to
\ket{\alpha} \ket{a \alpha+\beta}$ (again ``$a$'' stand for any power of $a$).
We write (everything $\bmod N$):
$$
\alpha ~a \quad = \quad (\alpha_0 + 2 \alpha_1 + \dots) ~a \quad = \quad
\alpha_0 ~a + \alpha_1 ~(2 a) + \alpha_2 ~(2^2 a) + \dots
$$
The numbers $2^j a$ (really we will have $2^j a^{2^i}$) can again be
precomputed and reduced $\bmod N$. Thus everything is decomposed into
$n$ controlled modular additions of precomputed integers in the range
$0 \dots N-1$. Each addition is conditioned on a different qubit of
the first (``$\alpha$-'') register.

We can leave the ``accumulation register'' in the coset representation during
all these $n$ additions. Under the plausible assumptions that the precomputed
numbers on average have a size of about $N/2$ and that on average we add up
only about half of these numbers, the sum will be around $n/2 \cdot N/2$.
I think that by subtracting (from the accumulation register) a multiple of $N$
close to this, we can improve the fidelity (thus reduce the error of the
modular addition technique).

So for a modular multiplication step $\ket{\alpha,\beta} \to
\ket{\alpha,\beta+a \alpha}$ we would leave the second register through all
$n$ modular additions in the ``coset'' representation. Usually we would
imagine that the first (``control'') register would be in the usual
representation, but actually we can also leave it in the coset representation.
Essentially this means that instead of multiplying with $\alpha$ we may
multiply with some $\alpha+x N$, but modulo $N$ this doesn't make any
difference. A disadvantage is that we now would have to carry out some $O(\log
n)$ more modular addition steps (per modular multiplication) and also that the
first register needs this many more qubits.

Thus in practise we may prefer to switch the first register back to
the usual representation while it acts as a control register. But if
we don't, the layout of Shor's algorithm becomes quite simple. We have
to convert the 2 registers to the ``coset representation'' only at the
beginning, and later no switching back will be needed. Each
multiplication step simply consists of a sequence of $n+O(\log n)$
regular additions. The only thing that doesn't look simple are the
``strange'' precomputed numbers we have to add... :-) Finally note
that while modular multiplication can be carried out with both
registers in the ``coset representation'', this wouldn't work as well
for an addition of the form $\ket{\alpha,\beta} \to
\ket{\alpha,\beta+\alpha}$ as then the fidelity loss would be large,
at least for the present scheme.

Note that the additions we do, naturally come out to be modulo the
size $2^{\#qubits}$ of the register, independently of whether we use a
classical reversible method or Draper's Fourier-addition. This is no
problem for the approximative modular addition described here. We can
choose the parameters such that a modular reduction never occurs, but
even if it does, the error per modular addition is still small.

If we use Draper's Fourier addition, we can leave the second (``target'')
register Fourier transformed for a whole sequence of additions. This allows
for a lot of parallelisation, as each addition can be carried out in a single
time step. (I think approximative classical reversible techniques exist which
allow a similar parallelisation of addition and also don't need auxiliary
work qubits. Here ``approximative'' would mean that the circuit works
correctly for all but a few inputs, which should be good enough for Shor's
algorithm. For such techniques see e.g. my 1998 work on implementing Shor's
algorithm.)

\subsection{Error estimate}

The question is how large we have to make the ``equal coset
superposition'' $\sum_x \ket{b+xN}$ (with $x=0\dots x_{max}-1$) to get
a good enough approximation for Shor's algorithm. Roughly we can argue
that in each modular addition we loose about $1/x_{max}$ in
fidelity. In order to keep the overall fidelity loss of the $4n^2$
additions at, say, below 1\%, we see that $x_{max} \approx 1000 n^2$
should be more that enough. This corresponds to using some $2\log
n+10$ additional qubits for the coset representation of a register.
Note that this ``analysis'' is not rigorous as it assumes that losses
in fidelity simply add up. While in general errors can behave worse
than that, I think that in the present case the assumption is
correct. (A rather easy ``worst case'' analysis shows that an
$x_{max}$ on the order of $n^4$ is provably enough.)

\subsection{Related work}

In 2000 Hales and Hallgren \cite{hallgren} have published an improved
approximate technique to carry out the quantum Fourier transform for
any order, e.g. for a large prime.  Their technique is simpler, faster
and uses fewer qubits than Kitaev's earlier method. One might think of
using their technique to carry out a quantum Fourier transform of
order $N$ on the register to which we have to add numbers modulo
$N$. Then Draper's Fourier addition technique could be used to
directly do additions modulo $N$. Actually the ``coset superposition''
technique which I propose is very similar to this. Hales and Hallgren
also first carry out the conversion $\ket{b} \to \sum_x \ket{b+x N}$
and then Fourier transform the whole register (thus modulo a power of
2).  Because they have Fourier transformed a ``periodic'' state, they
get peaks which they then ``extract'' to get the final result. If in
conjunction with my ``coset superpositions'' we also use Draper's
Fourier addition (modulo some $2^{n'}$), we thus essentially do the
same as Hales and Hallgren, except their last step, the ``extraction''
of the Fourier peaks, which fortunately turns out to not being necessary.

Also I understand that John Watrous \cite{watrous} has been using uniform
superpositions of subgroups (and cosets) in his work on quantum algorithms for
solvable groups. Thus he also used coset superpositions to represent elements
of the factor group (and probably also to carry out factor group operations on
them). In our case the overall group are the integers, the (normal) subgroup
are the multiples of $N$. The factor group who's elements we want to represent
is $Z_N \cong Z/(N Z)$. We now represent these elements by superpositions over
the cosets of the form $b+x N$. A problem in our case is that we can do things
only approximatively as the integers and the cosets are infinite sets.

\subsection{Some wishful thinking...}

As I have pointed out it has already been shown that much of the
computation in Shor's algorithm can be carried out classically (as
pre- and post-processing). It would be nice if the quantum part of
Shor's algorithm could be further reduced (at the expense of a
``reasonable'' amount of additional classical computation). E.g. it is
not clear whether maybe a modular multiplication step could not be
simplified. It is not even clear whether we could not carry it out
with only $O(n)$ quantum gates and maybe also with only one $n$-qubit
register, although I doubt that this is possible.

It would also be nice to find better quantum techniques for other
modular arithmetic operations. Namely a while ago I have
unsuccessfully tried to think about how ``modular inversion'' $x \to
x^{-1} (\bmod N)$ could be done more elegantly (maybe using Fourier
transforms or the like) than the classical (and ``classic'' :-) )
technique. In 2003 with John Proos in a work on elliptic curves we
used a cumbersome reversible implementation of Euclid's algorithm to
do that.

Also note that further simplifications of Shor's algorithm might lead
to some insights (into quantum computation). E.g. if the quantum part
could be reduced to some more natural operations. Also nice would be
if we could do without any pure qubits and projective measurements,
like the trace estimation problem of Knill and Laflamme in their
``power of one qubit'' paper. Another possibly practically useful
advance would be to ``break up'' Shor's algorithm into several smaller
(quantumly) independent quantum parts, but again I don't see how this
could be achieved. One can also investigate the possibility to replace
the modular multiplications by numbers of the form $a^{2^i} \bmod N$
by multiplications with other powers of $a$ or with any other suitably
chosen numbers.

\section{Shor's algorithm with $1.5n$ qubits (maybe less)}

\subsection{Modular multiplication with smaller factors}

Usually we have to multiply (mod $N$) a quantum register with a fixed
$n$-bit number. Here I show that if this classical factor is shorter,
we can accordingly save qubits. Later I will show how we can do Shor's
algorithm with such shorter factors, namely how we can replace a
single $n$-bit factor by two $n/2$-bit factors.

So say the classical factor ``$a$'' we want to multiply with has only $n'$
bits with $n'<n$, e.g. $n'=n/2$. We want to do $\ket{\alpha} \to \ket{a \cdot
  \alpha \bmod N}$ while using fewer work qubits than usually. I propose the
following sequence of three steps:
%
%$$
%\alpha ~~\to~~ a \cdot \alpha ~~\to~~ a \cdot \alpha \bmod N (=a \cdot \alpha
%- q \cdot N)~ ,~q ~~\to~~ a \cdot \alpha \bmod N 
%$$
%
\begin{eqnarray*}
\lefteqn{\alpha ~~ \mbox{($n$ bits)} \quad \to 
\quad a \cdot \alpha ~~ \mbox{($n+n'$ bits)} \quad \to} \\ 
&& a \cdot \alpha \bmod N (=a \cdot \alpha - q \cdot N)~ ,q ~~
\mbox{($n+n'$ bits)} \quad \to \quad a \cdot \alpha \bmod N 
~~\mbox{($n$ bits)}
\end{eqnarray*}
The first step is a usual multiplication (not modular). In the second
step we divide by $N$, getting the remainder and the (integer)
quotient $q$. In the last step we ``uncompute'' the quotient.

The first step (normal multiplication) is rather straight-forward. It is a
sequence of controlled additions, conditioned on bits of $\alpha$ after each
of which the controlling bit is uncomputed. Note that this way the total
number of qubits is always at most $n+n'$. We can go through the bits of
$\alpha$ both ways but it's a bit easier if we go from most significant to
least significant as then uncomputing the bits of $\alpha$ is
easiest. (E.g. when $a$ is odd the lowest bit of the evolving sum is simply
equal to the bit we need to uncompute.)

The second step is the reduction modulo $N$. It is a usual division consisting
of a sequence of subtractions of $2^i \cdot N$ for $i=n',n'-1, \dots ,0$. For
each subtraction one bit of the quotient $q$ is computed, from most to least
significant, each bit indicating whether the corresponding subtraction was
done or not.

The third step (uncomputing of $q$) is less straight forward. I show how the
reverse can be done, thus computing $q$ from the remainder $a \alpha \bmod N =
a \alpha -q N$. First consider computing the remainder of this modulo $a$: $(a
\alpha - q N) \bmod a = -q N \bmod a$. Computing the remainder of a quantum
register modulo a fixed number is indeed not hard to do. We need an
``accumulation'' register into which we can add numbers modulo $a$. For each
qubit of the original quantum register we then add, conditioned on this qubit,
the (classically precomputed) number $2^i \bmod a$. To directly obtain $q$
(instead of $-q N \bmod a$) we now simply replace these numbers with $2^i
(-N)^{-1} \bmod a$.

Thus in total we need only $n+n'$ qubits for modular multiplication with
factors of size $n'$ (with $n'<n$).

\subsection{Getting multiplication factors of size $n/2$}

Usually in Shor's algorithm we have to multiply by $n$-bit factors, namely by
the numbers $a^{2^i} \bmod N$. If we could factor these numbers into smaller
numbers, we could accordingly save qubits. Here I show how we can write these
numbers as quotients ($\bmod N$) of two half size numbers. (Note that dividing
by such a number is simply the reverse of multiplying.) Say the original
factor was $a$. Now consider the extended Euclidean algorithm one would use to
compute $a^{-1} \bmod N$. In this algorithm we get a sequence of integer
linear combinations of $a$ and $N$ of the form $r \cdot a + k \cdot N =
r'$. In the course of the algorithm the integer $r$ gets larger while $r'$
gets smaller, while always $|r| \cdot |r'| < N$. By stopping the algorithm in
the middle we get two numbers $r,r'$ both of size about $\sqrt{N}$, with
$a=r'/r \bmod N$. (Note that the integer coefficients $a$ and $k$ can be
negative, but $r'$ is usually taken to remain positive.)

Actually it's not hard to show that there appears always a linear
combination (and sometimes two) in Euclid's algorithm where (the
absolute values of) both $r$ and $r'$ are smaller than $\sqrt{N}$.

In summary this gives us a factoring algorithm with about $1.5n$
qubits. In a more careful count I arrive at $1.5n+2$ qubits (when $n$
is odd, round up). For this I used a modular addition circuit which
uses only 1 work qubit and of course also Draper's Fourier addition
technique to save qubits. (By the way, I wonder whether there is maybe
a general way to get rid of a single work qubit, e.g. at the expense of
increasing the number of gates by some factor, say 8 ...?)

\subsubsection{Even smaller factors?}

If we could somehow write our $n$-bit factors $a$ as a product (or quotient)
of even smaller factors we could further save qubits. One idea might be to try
to (further) factor the $n/2$-bit numbers obtained above. A full factorisation
would be hard. (But still less hard than factoring $n$-bit numbers...
actually one idea is to use the quantum algorithm to do this. This wouldn't
increase the number of qubits, but would increase the quantum running time
quite a bit). Easier would be to just look for small primefactors. At any rate
for each step we would need to have a choice of numbers which we could try to
(partially) factor. Indeed given the linear combinations in the middle of
Euclid's algorithm one can find many $r,r'$ pairs with sizes of about $n/2$
bits. Also the $n$-bit factors $a$ need not be the numbers usually
taken. (E.g. one can think of a scheme whereby one is allowed to use
essentially arbitrary products of the numbers $a^{2^i}\bmod N$ for small
ranges of $i$.) Note that we can't hope to find enough fully ``smooth''
numbers (numbers with only small primefactors), at least not for $n$-bit
numbers, as otherwise one could use this for efficient classical factoring...
but I think that maybe (classical) factorisation into small enough factors is
possible to be able to run Shor's algorithm with only, say, $1.1n$ qubits.

\section{Remarks}

\subsection{Useful for future real implementations?}

Note that here we haven't tried to find a ``best'' (most efficient) circuit
for future actual implementations. The tradeoffs one would like to make for
physical realisations are not clear, e.g. how many more qubits one is willing
to use to speed the algorithm up. Also in reality one probably needs many more
qubits for error correction and the spatial arrangement and connections
between qubits are an issue. Still, maybe some of the techniques worked out
for abstract circuits may one day be useful...

\subsection{Optimisation of reversible and quantum circuits}

There seem to be quite some opportunities for the optimisation of the
quantum part of a circuit (at the expense of a ``reasonable'' amount
of classical computation). If we restrict just to classical reversible
circuits the task can be formulated as finding (efficiently
computable) short circuits. Note that the possibility to shift work to
the classical (fixed) part gives us extra opportunities for
optimisation. Also the possibility of doing things only approximately
(e.g. wrong for a small fraction of the computational basis states)
allows for extra possibilities.

\subsection{Conclusions}

We have sketched three results. One is that even fewer qubits need to be
initialised for factoring than known so far. Then a non-classical modular
addition technique with ``coset superpositions'' was proposed which may be of
use for parallelising computations. Finally I have shown how instead of the
usual $2n+\dots$ qubits, we can run Shor's algorithm with only  $\lceil 1.5n
\rceil+2$ qubits. It seems likely that this can be further reduced, at least
somewhat.

\end{document}